\documentstyle[12pt,
epsfig]{article}
\relax

\newcommand{\phib}{\bar{\phi}}
\newcommand{\phit}{\tilde{\phi}}
\newcommand{\mt}{\tilde{m}}
\newcommand{\nt}{\tilde{n}}
\newcommand{\Nt}{\tilde{N}}
\newcommand{\Bt}{\tilde{B}}
\newcommand{\Rt}{\tilde{R}}
\newcommand{\rt}{\tilde{r}}
\newcommand{\mut}{\tilde{\mu}}

\newcommand{\rhot}{\tilde \rho}

\newcommand{\pt}{\tilde{p}}
\newcommand{\Ut}{\tilde{U}}

\newcommand{\sx}{\sigma}

\newcommand{\be}{\begin{equation}}
\newcommand{\ee}{\end{equation}}

\def \gta {\mathrel{\vcenter
     {\hbox{$>$}\nointerlineskip\hbox{$\sim$}}}}

\begin{document}

\begin{center}
{ \Large \bf
Neutrino Lumps in Quintessence Cosmology} 
\\
\vspace{1.5cm}
{\Large 
N. Brouzakis$^{1}$, N. Tetradis$^{1}$ and C. Wetterich$^2$ 
} 
\vspace{0.5cm}
\\
{\it
$^1$
Department of Physics, University of Athens, 
\\
Zographou GR-15784, Athens, Greece
\\
$^2$
Institut f\"ur Theoretische Physik, Universit\"at Heidelberg, 
\\
Philosophenweg 16, 
D-69120 Heidelberg, Germany} 
\end{center}
\vspace{3cm}
\abstract{
Neutrinos interacting with the quintessence field can trigger the accelerated expansion 
of the Universe. 
In such models with a growing neutrino mass
the homogeneous cosmological solution 
is often unstable to
perturbations. We present static, spherically symmetric solutions of the 
Einstein equations in the same models. They describe astophysical objects composed of
neutrinos, held together by gravity and the attractive force mediated by the quintessence field. 
We discuss their characteristics as a function of the present neutrino mass.
We suggest that these objects are the likely outcome of the growth of
cosmological perturbations.  
}

\newpage
The mechanism responsible for the onset of the accelerating phase in quintessence 
cosmology remains undetermined. Explaining 
the emergence of an accelerating phase in recent cosmological times
constitutes one of the most difficult challenges of quintessence models - the coincidence problem. 
A possible trigger for the acceleration
has been proposed recently \cite{amendwet,wet}, arising through
the interaction of the quintessence field with
a matter component whose mass grows with time. 
This matter component may be identified with neutrinos
\cite{amendwet,wet,mavan}. In the proposed scenario the neutrinos remain essentially massless
until recent times. When their mass eventually grows close to its present value, their 
interaction with the quintessence field (the cosmon) almost stops its evolution.
The potential energy of the cosmon becomes 
the dominant contribution to the energy density of the Universe. Cosmological 
acceleration ensues.

For the coupled neutrino-cosmon fluid the squared sound speed $c^2_s$ may become negative -
a signal of instability \cite{mavan}.
Indeed, the sign of $c^2_s$ oscillates in the accelerating phase for one of the proposed models
\cite{wet}. A natural interpretation of this instability is
that the Universe becomes inhomogeneous with the neutrinos 
forming denser structures. 
Within the linear approximation the neutrino fluctuations can be followed in these
models until a redshift around one, when the neutrino overdensities become nonlinear
\cite{linear}. One suspects that some form of subsequent 
collapse of these fluctuations will result into bound neutrino lumps.
In this letter we present static, spherically symmetric
solutions of the Einstein equations 
that describe such structures and study their characteristics. 
Astrophysical objects composed of neutrinos have also been studied in  
\cite{stephenson,bilic}.

We assume that the energy density of the Universe involves a gas of weakly interacting 
particles (neutrinos). The mass $m$ of the particles depends on
the value of a slowly varying cosmon field $\phi$ \cite{wetcosmon}. 
For the field equation 
\be
\frac{1}{\sqrt{-g}}\frac{\partial}{\partial x^\mu}
\left(\sqrt{-g}\,\,g^{\mu\nu}\frac{\partial \phi}{\partial x^\nu}
\right)=
\frac{dU}{d\phi}-\frac{1}{m}\frac{d m(\phi(x))}{d\phi}\,\, T^\mu_{~\mu}.
\label{eomphi} \ee
we approximate the neutrino energy-momentum tensor as $T^\mu_{~\nu}={\rm diag} (-\rho,p,p,p)$. 
The cosmology of \cite{amendwet,wet} 
also assumes the presence of another gas of particles (dark matter) whose mass is
independent of $\phi$.

We consider stationary, spherically symmetric configurations, with metric 
\be
ds^2=-B(r)dt^2+r^2(d\theta^2+\sin^2\theta\, d\varphi^2)+A(r)dr^2.
\label{metric} \ee
For the neutrinos we assume a Fermi-Dirac distribution, with locally varying density - 
the Thomas-Fermi approximation.
The local chemical potential satisfies 
$\mu(r)=\mu_0/\sqrt{B(r)}$ \cite{bilic,tetradis}.
Stable
configurations are prevented from collapsing by the pressure generated through the 
exclusion principle. 
We concentrate on vanishing temperature of the neutrino gas.
We do not expect qualitative changes of our solution 
for a non-zero temperature.
For simplicity we consider one neutrino species, with the generalization to 
degenerate neutrino masses being straightforward.

We parametrize the particle mass by a dimensionless function $\mt$, defined according to
$m(\phi)= \sx \mt\left[{(\phi-\phib)}/{M}\right]$,
with $\sx$ an arbitrary energy scale and $M=(16\pi G)^{-1/2}\simeq 1.72\times 10^{18}$ GeV.
Here $\phib$ is a fixed reference value, close to the present value of the quintessence field.
Hence, $\sx$ is of the order of the present neutrino mass, in the eV range or somewhat below.
For concreteness, we consider a cosmon potential of the form
$U(\phi)=C \sx^4\exp\left[-a{(\phi-\phib)}/{M} \right]$
with $a={\cal O}(1)$. However, the effect of the potential on our solutions 
is negligible. For this reason, the predicted astrophysical objects 
are largely independent of the form of the potential, and depend mainly on the 
interaction between dark energy and neutrinos.
The present cosmological value of $\phi$ 
is given by the requirement that $U(\phi)$ 
constitute about 3/4 of the 
critical energy density $U(\phi)\simeq 10^{-11}$ (eV)$^{4}$. The cosmological value
of $\phi$ is taken as the asymptotic value $\phi_{as}$ of our local solutions for large $r$,
obeying
$(\phi_{as}-\phib)/M=\phit_{as} \simeq ({1}/{a})
\left[ 25.3+\ln C +4 \ln \left( {\sx}/{{\rm eV}}\right)\right]$.

The equations of motion become more transparent if we define the
dimensionless variables 
$\phit={(\phi-\phib)}/{M}$
and $\rt= {\sx^2r}/{M}$.
All other dimensionful quantities are multiplied with appropriate
powers of $\sx$, in order to form dimensionless quantities denoted as
tilded. We use
$\Bt = {B}/{\mut^2_0}={B\sx^2}/{\mu_0^2}$ and 
$\mut(\rt)={1}/{\sqrt{\Bt(\rt)}}$.
We define the radius $\Rt$ of the compact object by the value of $\rt$ 
at which the fermionic density becomes negligible. 
The physical radius is 
$${R}/{{\rm Mpc}}\simeq 1.1\times 10^{-2}\left(\sx/{\rm eV}\right)^{-2} \Rt.$$
The mass of the object is given by its Schwarzschild radius $\Rt_s$. For
$\rt \to \infty$ we have $B=1/A=1-\Rt_s/\rt$. In units of the solar mass, the mass of the neutrino lump
is $${M_{tot}}/{M_\odot}\simeq 1.2 \times 10^{17} \left(\sx/{\rm eV}\right)^{-2} \Rt_s.$$
Another important characteristic is the total 
neutrino number, which we assume to be conserved. It is
$$N\simeq 5.1\times 10^{81} \left(\sx/{\rm eV} \right)^{-3} \Nt,$$ 
with  
$$\Nt=\int_0^\infty 4 \pi \rt^2 \nt \sqrt{A}  d\rt.$$

The field equations read \cite{tetradis}
\begin{eqnarray}
\phit''+\left(\frac{2}{\rt}-\frac{A'}{2A}+\frac{\Bt'}{2\Bt} \right)
\phit'&=&
A\left[ \frac{d\Ut}{d\phit}+\frac{1}{\mt}\frac{d\mt}{d\phit}\,\,(\rhot-3\pt)\right]
=-A \frac{d\left( \pt-\Ut \right)}{d\phit},
\label{eqphid} \\
\frac{1}{\rt^2}\frac{1}{A}-\frac{1}{\rt^2}-\frac{1}{\rt}\frac{A'}{A^2}
&=&\frac{1}{2}\left[ -\frac{1}{2A}\phit'^2-\Ut(\phit)-\rhot\right],
\nonumber \\
\frac{1}{\rt^2}\frac{1}{A}-\frac{1}{\rt^2}+\frac{1}{\rt}
\frac{\Bt'}{\Bt A}
&=&\frac{1}{2}\left[ \frac{1}{2A}\phit'^2-\Ut(\phit)+\pt\right],
\nonumber 
\end{eqnarray}
where a prime denotes a derivative with respect to $\rt$.
We also have
\begin{eqnarray}
\nt&=&\frac{1}{3\pi^2} \left(\mut^2-\mt^2  \right)^{3/2},
\label{nd} \\
\pt&=&\frac{1}{24\pi^2}
\left[\mut \sqrt{\mut^2-\mt^2}\left(2\mut^2-5\mt^2 \right) 
+3\mt^4\ln\left(
\frac{\mut+\sqrt{\mut^2-\mt^2}}{\mt} \right)
\right],
\nonumber \\
\rhot&=&\frac{1}{8\pi^2}
\left[\mut \sqrt{\mut^2-\mt^2}\left(2\mut^2-\mt^2 \right) 
-\mt^4\ln\left(
\frac{\mut+\sqrt{\mut^2-\mt^2}}{\mt} \right)
\right],
\nonumber 
\end{eqnarray}
for $\mut\geq \mt$, and $\nt=\rhot=\pt=0$ for $\mut< \mt$.
Finally,
$\Ut(\phit)=C \exp\left(-a \phit \right)$.

\begin{figure}[t]
\begin{center}
\includegraphics[clip,width=0.75\linewidth]{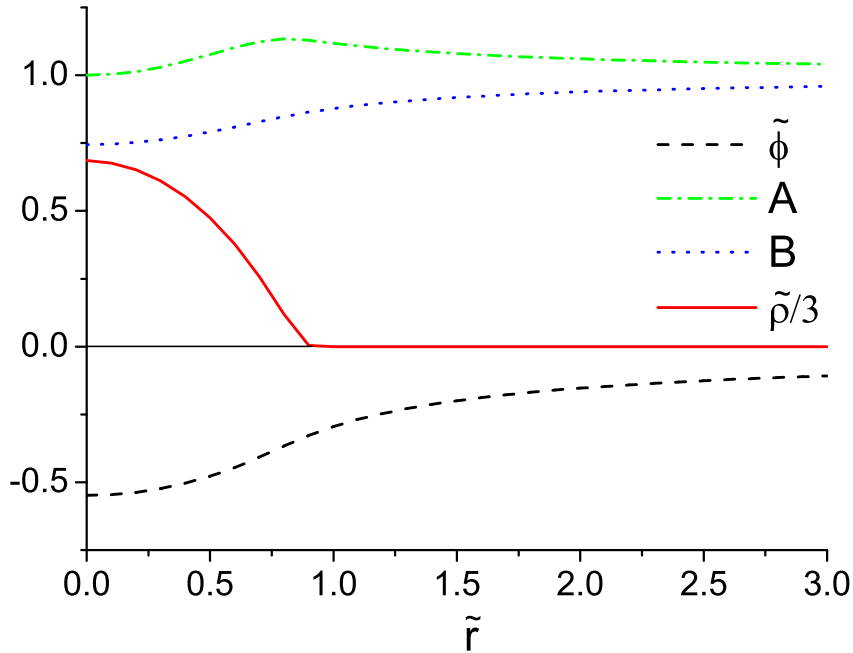}
\caption{Radial dependence of $\phit$, $\rhot$, $A$ and $B$ for a neutrino lump.}
 \label{fig1}
 \end{center}
  \end{figure}

We need four 
initial conditions for the system of equations (\ref{eqphid}).
Two of them are imposed by the regularity of the 
solution at $\rt=0$: $\phit'(0)=0$, $A(0)=1$.
The value of $\Bt(0)$ is the only free integration constant. Since
$AB(r\to\infty)=1$ one has
$A\Bt(\rt\to\infty)=(\mu_0/\sx)^{-2}$. As a result, the choice of
$\Bt(0)$ determines the chemical potential and, therefore, the total number of
neutrinos in the lump.
Finally, $\phit(0)$ must be chosen so that 
$\phit(\rt\to\infty)$ reproduces correctly the present
value $\phit_{as}$ of the cosmological solution. 
(We assume that the time scale of the cosmological 
solution is very large and neglect the time dependence of 
$\phit(\rt\to\infty)$.)

We consider two types of models, distinguished by the dependence of the particle mass on
the field: \\
Model I assumes 
$\mt(\phit)=- {1}/{\phit}$ \cite{wet},
with the field $\phit$ taking negative values.
\\
Model II assumes
$\mt(\phit)=\exp \left(-b \phit \right)$ \cite{amendwet},
with $b < 0$.
(Notice that $a=\alpha/\sqrt{2}$, $b=\beta/\sqrt{2}$ in comparison to \cite{amendwet}, where a different
convention for $M$ is used.)

In both cases we are interested in values of the field near $\phit=0$. 
For model II we can choose $\phib$ such that $\phit_{as}=0$, implying that $\sx=m_\nu(t_0)$ equals the
present neutrino mass. One infers for the quintessence potential
$\ln C=-25.3-4\ln \left(\sx/{\rm eV} \right)$.
The parameter $a$ is fixed by requiring that during the early stages of the
cosmological evolution the
dark energy be subleading and track the radiation or the dark matter. 
During the radiation and matter dominated epochs, the dark energy follows
a ``tracker'' solution with a constant density parameter
$\Omega_{h,early}=n/(2 a^2)$, where $n=3\,(4)$ for matter (radiation) \cite{amendwet}.
Observations require $a$ to be large, typically $a \gta 7$ \cite{doran}.
We use $a=7$ in the following.
The future of our Universe is described by a different attractor, for which the
dark energy dominates. Our present era coincides with the transition between
the two cosmic attractors. The influence of the neutrinos on the evolution of the
cosmon field is determined by the second term in the r.h.s. 
of eq. (\ref{eomphi}). Demanding that today this term be equal to the first term, that
arises from the potential, fixes 
the present neutrino fraction to the value
$\Omega_\nu(t_0)=-(b/a)\Omega_h(t_0)$ \cite{amendwet}.
For a realistic cosmology with present dark energy fraction $\Omega_h(t_0) \simeq 3/4$ one
has to adjust $b$ to the neutrino mass. 
For one dominant neutrino species we have
$b=-a (36 \,{\rm eV}/m_\nu(t_0))$.
For model I we need to know how close $\phit_{as}$ is to zero, with 
$\sx=-\phit_{as} m_\nu(t_0)$. As compared to model II, we have now an effective
$\phit$-dependent $b(\phit)=-1/\phit$, which results in the condition
$\phit_{as}=-(1/a)m_\nu(t_0)/(36\,{\rm eV})$ or $\sx=(1/a)m_\nu^2(t_0)/(36\,{\rm eV})$ .

In fig. \ref{fig1} we present a typical solution describing a static astrophysical object
in model I.
The chemical potential has the value $\mut_0\simeq 2.9$.
The scalar field becomes more negative
near the center of the solution, so that the neutrinos become 
lighter there. The asymptotic value is $\phit_{as}=-0.02$, which corresponds to 
$m_\nu(t_0)\simeq 5$ eV. 
The pressure and density of the fermionic gas 
vanish for $\rt \geq \Rt \simeq 0.91$. The mass of the object can be
deduced from the asymptotic form of $A$ or $B$ for $\rt \to \infty$. 
We find $\Rt_s \simeq 0.12$. The total fermionic number is $\Nt\simeq 0.88$.
The form of the solutions in model II is similar to the one depicted in fig. \ref{fig1}.

\begin{figure}[t]
\begin{center}
\includegraphics[clip,width=0.75\linewidth]{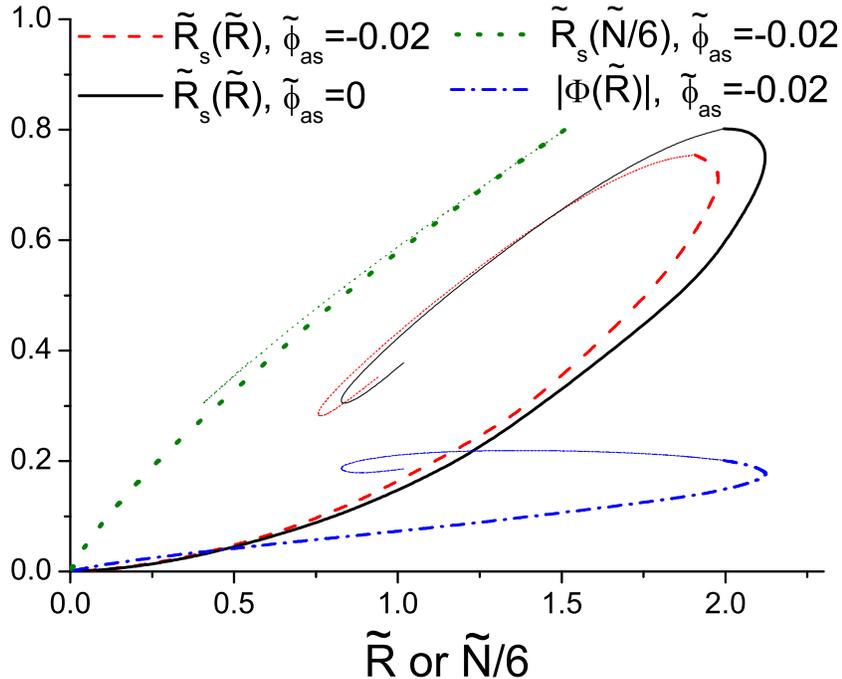}
\caption{Mass vs. size for neutrino lumps in model I.}
 \label{fig2}
 \end{center}
  \end{figure}

The variation of the chemical potential results in a whole class of solutions, depicted by
the solid line in fig. \ref{fig2}. We display the dimensionless Schwarzschild radius $\Rt_s$
as a function of the
dimensionless radius of the object $\Rt$. There is a maximal value for the mass, denoted
by the end of the thick line. The continuation of the curve has the form of a spiral and is depicted
by a thinner line. This branch is unstable to perturbations that can lead to gravitational 
collapse \cite{tdlee1}. 
In order to demonstrate this fact, we 
plot in the same figure $\Rt_s$ as a function of $\Nt/6$ (dotted line). 
This curve has two branches. The one depicted by a thinner line
corresponds to the thinner line of the curve $\Rt_s(\Rt)$. There are two
possible values of $\Rt_s$ that correspond to the same value of the total neutrino number $\Nt$. 
The value on the thinner line has
a larger value of $\Rt_s$ and results in a larger mass. The corresponding configuration is unstable towards
one with the same $\Nt$ located on the thicker line. 
The characteristics of the solutions depend only very mildly on the value of $\phit_{as}$, as demonstrated 
by the comparison of the solid and dashed curves. All the values of $m_\nu(t_0)$ in the range [0,5] eV
correspond to $\phit_{as}$ in the range [-0.02,0]. The respective $\Rt_s(\Rt)$ curves lie between 
the solid and dashed curves of fig. \ref{fig2}.

A striking feature is the existence of neutrino lumps with arbitrarily small mass. They correspond
to the lower left corner of the figure, where  both $\Rt$ and $\Rt_s$ vanish.
For such objects the contribution from gravity is negligible and their existence 
is a consequence of the attractive force mediated by the scalar field. 
Such configurations are not
generic, but depend crucially on the assumed form of $\mt(\phit)$. A completely different form of
solutions appears in model II.
In fig. \ref{fig2} we also depict
the gravitational potential $\Phi(\rt)=-\Rt_s/(2\rt)$ at a distance $\rt=\Rt$
equal to the radius of the astrophysical object.

\begin{figure}[t]
\begin{center}
\includegraphics[clip,width=0.75\linewidth]{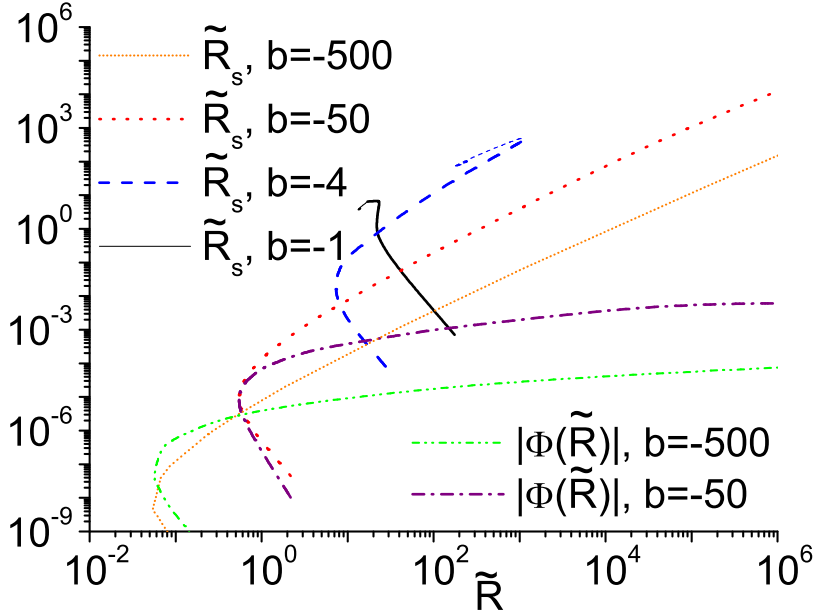}
\caption{Same for model II.}
 \label{fig3}
 \end{center}
  \end{figure}

The function $\Rt_s(\Rt)$ in model II displays a different behaviour. 
In fig. \ref{fig3} we plot this function 
for four different values of $b$, namely $b=-500,-50,-4,-1$. Realistic neutrino masses
correspond to large, negative $b$.
For $b=-1$ (solid line) there is a maximal value for the mass of the astrophysical objects and
a branch of unstable solutions. The maximal value of $\Rt_s$ is comparable for model I and 
model II with $b=-1$, even though the corresponding radius is larger by an order of
magnitude in the second case. For $b=-4$ (dashed line)
the maximal value of $\Rt_s$ and the corresponding $\Rt$
increase by roughly two orders of magnitude. 

The crucial qualitative difference with model I concerns the form of the solutions with low values
of $\Rt_s$. In model I for $\Rt_s \to 0$ we have $\Rt\to 0$, while in model II we have $\Rt\to \infty$.
The attractive interaction mediated by the scalar field in model II
is not sufficiently strong to lead to bound objects with a small
fermion number. Gravity must play a role for compact objects to exist. As $|b|$ increases the
dependence of $\mt$ on $\phit$ becomes more pronounced. The 
effective neutrino mass in the interior of a compact object can become smaller without a
large variation of $\phit$ (and a significant energy cost through the field derivative term). 
This has two significant effects: a) Objects with smaller $\Nt$ and $\Rt$ can exist. As a result the
bending of the curve $\Rt_s(\Rt)$ for low $\Rt_s$ takes place for smaller $\Rt$. b) The configurations 
that are gravitationally unstable (indicated by the spiral in the upper part of the curve) are shifted
toward larger values of $\Rt_s$. The reason is that the neutrinos are essentially massless in the
interior of of such configurations, carrying only kinetic energy. This makes the gravitional collapse 
difficult.

The curves $\Rt_s(\Rt)$ in model II with
$b=-500$  and $-50$ are also
depicted in fig. \ref{fig3}. We have not managed to determine numerically a maximal value of $\Rt_s$, as
objects with huge values of $\Rt_s$, $\Rt$ (larger by more than twenty 
orders of magnitude than the ones depicted) 
are possible. 
For comparison we note than in model I we have a maximal value $(\Rt_s)_{max}=0.80$ with a corresponding
radius $(\Rt)_{max}=2.0$.
In fig. \ref{fig3} we observe minimal values of the radius, $(\Rt)_{min}=0.54$ for $b=-50$ and
$(\Rt)_{min}=0.054$ for $b=-500$.
The corresponding values of the Schwarzschild radius are $(\Rt_s)_{min}=1.1\times 10^{-5}$ 
and $(\Rt_s)_{min} =1.1 \times 10^{-8}$, respectively.
It is apparent that for small $\Rt_s$ we have the scaling behaviour $\Rt \sim b^{-1}$,
$\Rt_s\sim b^{-3}$. This can be understood by noticing that in the limit $A',B'\to 0$, $A\to 1$,
and for negligible $d\Ut/d\phit$, the factors of $b$ 
in eq. (\ref{eqphid}) can be eliminated through the redefinitions $b \phit\to \phit$, $b\rt \to \rt$.
In fig. \ref{fig3} 
we also depict the surface gravitational potential
$\Phi(\Rt)=-\Rt_s/(2\Rt)$ as a function of $\Rt$ for the cases $b=-500$ and $-50$.

\begin{figure}
\begin{center}
\includegraphics[clip,width=0.71\linewidth]{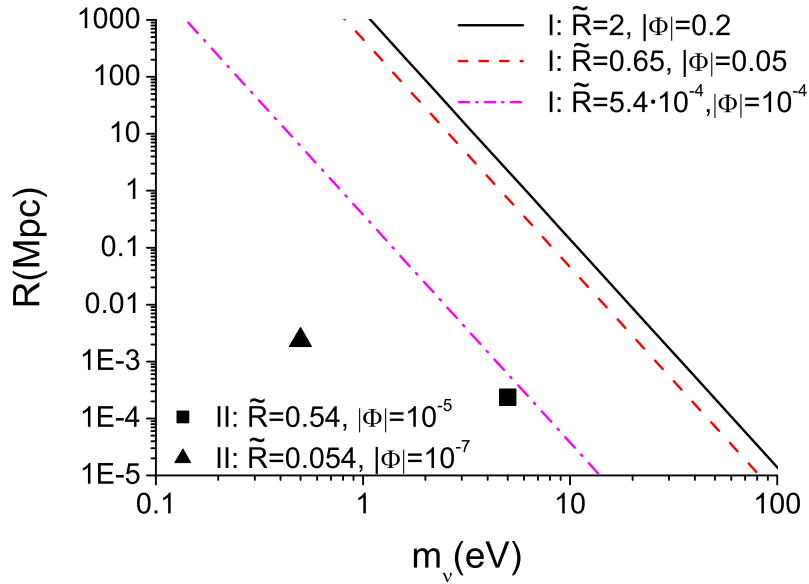}
\caption{Size of neutrino lumps as a function of the neutrino mass.}
 \label{fig4}
 \end{center}
  \end{figure}

In fig. \ref{fig4} we display the size $R$ of the astrophysical objects as 
a function of the present neutrino mass $m_\nu\equiv m_\nu(t_0)$.
Restoring physical units requires the scale $\sx$. We use for model I $a=7$ or
$(\sx/{\rm eV})^{1/2}\simeq 0.063
(m_\nu/{\rm eV})$. The function $\Rt_s(\Rt)$ has a very mild dependence on 
$\phit_{as}$ for $0\leq \phit_{as}\leq 0.02$ (see fig. \ref{fig2}). 
For given $\Rt$ the variation of $\sx$ (or equivalently $m_\nu$) produces a 
class of astrophysical
objects of variable physical size. They all generate the same surface gravitational
potential $\Phi=-\Rt_s/(2\Rt)$. 
In fig. \ref{fig4} we depict three such classes. The first two  
contain objects with strong gravitational potentials, while the last one contains objects that generate
weaker fields. Solutions with $\Rt \to 0$ produce curves parallel to
those in fig. \ref{fig4}, but located closer to the lower left corner.
In the same figure we also depict two solutions 
of model II. In this model the neutrino mass is uniquely determined by the value of $b$.
The two points in fig. \ref{fig4} correspond to the minimal values
of $\Rt$ for $b=-50$ and $-500$. These are $\Rt=0.54$ and $\Rt=0.054$, respectively.

Recently, a first investigation of the coupled fluctuations of dark matter, neutrinos,
baryons and the cosmon field has been performed for the models within the linear 
approximation \cite{linear}. For a specific model with a present average neutrino
mass of 2.1 eV, the neutrino fluctuations grow nonlinear at a redshift around one.
The typical size of these fluctuations is large, in the range of superclusters
and beyond. A further investigation of the fate of these neutrino lumps will have to
follow their collapse due to the scalar-mediated attractive 
interaction and gravity. This should generate the distribution of the integration
constants of the present solution, like the characteristic mass and size of
the lumps.

Our study demonstrates that the presence of instabilities in quintessence cosmologies with a variable
neutrino mass may have interesting astrophysical consequences. After a sufficiently long time, 
these instabilities may lead to the formation of stable bound neutrino lumps.
Their radius and mass within the family of allowed solutions (for given $m_\nu$) depend on the details
of the dynamical formation mechanism. Since in the models of \cite{amendwet,wet} the neutrinos
remain free streaming until a rather recent cosmological epoch (say, $z=5$), one may expect a large
typical size of the neutrino lumps (more than 100 Mpc). At the present stage of the investigations
it is not clear if such lumps have already decoupled from the cosmological expansion -
for this, the perturbations have to grow nonlinear - or if this will happen only in the future.
In the extreme case of an early formation of a population of lumps with subgalactic size, they could 
even play the role of dark matter. The detection of lumps could proceed directly through their
gravitational potential, or indirectly through their attraction for baryons. Quintessence cosmologies
may provide surprises for structures on very large scales.

{\it Acknowledgments}:
This work was supported by the research program
``Pythagoras II'' (grant 70-03-7992)
of the Greek Ministry of National Education, partially funded by the
European Union.



\begin{thebibliography}{999}


\bibitem{amendwet}
  L.~Amendola, M.~Baldi and C.~Wetterich,
  arXiv:0706.3064 [astro-ph].

\bibitem{wet}
  C.~Wetterich,
  Phys.\ Lett.\  B {\bf 655} (2007) 201
  [arXiv:0706.4427 [hep-ph]].

\bibitem{mavan}
  P.~Gu, X.~Wang and X.~Zhang,
  Phys.\ Rev.\  D {\bf 68} (2003) 087301
  [arXiv:hep-ph/0307148];
\\
  R.~Fardon, A.~E.~Nelson and N.~Weiner,
  JCAP {\bf 0410} (2004) 005
  [arXiv:astro-ph/0309800];
\\
  A.~W.~Brookfield, C.~van de Bruck, D.~F.~Mota and D.~Tocchini-Valentini,
  Phys.\ Rev.\ Lett.\  {\bf 96} (2006) 061301
  [arXiv:astro-ph/0503349];
\\
  N.~Afshordi, M.~Zaldarriaga and K.~Kohri,
  Phys.\ Rev.\  D {\bf 72} (2005) 065024
  [arXiv:astro-ph/0506663];
\\
  O.~E.~Bjaelde, A.~W.~Brookfield, C.~van de Bruck, S.~Hannestad, D.~F.~Mota, L.~Schrempp and D.~Tocchini-Valentini,
  JCAP {\bf 0801} (2008) 026
  [arXiv:0705.2018 [astro-ph]].
\\
  K.~Ichiki and Y.~Y.~Keum,
  arXiv:0705.2134 [astro-ph];
\\
  R.~Bean, E.~E.~Flanagan and M.~Trodden,
  arXiv:0709.1128 [astro-ph].

\bibitem{linear}
  D.~F.~Mota, V.~Pettorino, G.~Robbers and C.~Wetterich,
  Phys.\ Lett.\  B {\bf 663} (2008) 160
  [arXiv:0802.1515 [astro-ph]].

\bibitem{stephenson}
  G.~J.~.~Stephenson, J.~T.~Goldman and B.~H.~J.~McKellar,
  Int.\ J.\ Mod.\ Phys.\  A {\bf 13} (1998) 2765
  [arXiv:hep-ph/9603392].

\bibitem{bilic}
  N.~Bilic and R.~D.~Viollier,
  Gen.\ Rel.\ Grav.\  {\bf 31} (1999) 1105
  [arXiv:gr-qc/9903034];
\\
 N.~Bilic, R.~J.~Lindebaum, G.~B.~Tupper and R.~D.~Viollier,
  Phys.\ Lett.\  B {\bf 515} (2001) 105
  [arXiv:astro-ph/0106209].

\bibitem{wetcosmon}
 C.~Wetterich,
  Astron.\ Astrophys.\  {\bf 301} (1995) 321
  [arXiv:hep-th/9408025].

\bibitem{tetradis}
  N.~Tetradis,
  Phys.\ Lett.\  B {\bf 632} (2006) 463
  [arXiv:hep-ph/0507288];
\\
  N.~Brouzakis and N.~Tetradis,
  JCAP {\bf 0601} (2006) 004
  [arXiv:astro-ph/0509755].

\bibitem{doran}
  M.~Doran, G.~Robbers and C.~Wetterich,
  Phys.\ Rev.\  D {\bf 75} (2007) 023003
  [arXiv:astro-ph/0609814].

\bibitem{tdlee1}
  T.~D.~Lee and Y.~Pang,
  Phys.\ Rev.\  D {\bf 35} (1987) 3678.
 

\end{thebibliography}
\end{document}